\documentclass[aps,prx,twocolumn,superscriptaddress]{revtex4}
\usepackage{graphicx}
\usepackage{amssymb}
\usepackage{amsmath}
\usepackage{epsfig}
\usepackage{color}
\usepackage{float}
\usepackage[colorlinks=true,
			linkcolor=blue,
			urlcolor=cyan,
			citecolor=blue]{hyperref}
\usepackage{hyperref}
\usepackage{lipsum}		
\usepackage{ulem}
\begin{document}
\title{Removing grid structure in angle-resolved photoemission spectra via deep learning method}
\author{Junde Liu }
\affiliation{Beijing National Laboratory for Condensed Matter Physics and Institute of
Physics, Chinese Academy of Sciences, Beijing 100190, China}
\affiliation{University of Chinese Academy of Sciences, Beijing 100049, China}
\author{Dongchen Huang }
\affiliation{Beijing National Laboratory for Condensed Matter Physics and Institute of
Physics, Chinese Academy of Sciences, Beijing 100190, China}
\affiliation{University of Chinese Academy of Sciences, Beijing 100049, China}
\author{Yi-feng Yang}
\email[Corresponding author:]{yifeng@iphy.ac.cn}
\affiliation{Beijing National Laboratory for Condensed Matter Physics and Institute of
Physics, Chinese Academy of Sciences, Beijing 100190, China}
\affiliation{University of Chinese Academy of Sciences, Beijing 100049, China}
\affiliation{Songshan Lake Materials Laboratory, Dongguan, Guangdong 523808, China}
\author{Tian Qian}
\email[Corresponding author:]{tqian@iphy.ac.cn}
\affiliation{Beijing National Laboratory for Condensed Matter Physics and Institute of
Physics, Chinese Academy of Sciences, Beijing 100190, China}
\affiliation{University of Chinese Academy of Sciences, Beijing 100049, China}
\affiliation{Songshan Lake Materials Laboratory, Dongguan, Guangdong 523808, China}

\date{\today}
\begin{abstract}
    Spectroscopic data may often contain unwanted extrinsic signals. For example, in the angle-resolved photoemission spectroscopy (ARPES) experiment, a wire mesh is typically placed in front of the CCD to block stray photo-electrons but could cause a grid-like structure in the spectra during quick measurement mode. In the past, this structure was often removed using the mathematical Fourier filtering method by erasing the periodic structure. However, this method may lead to information loss and vacancies in the spectra because the grid structure is not strictly linearly superimposed. Here, we propose a deep learning method to overcome this problem effectively. Our method takes advantage of the self-correlation information within the spectra themselves and can greatly optimize the quality of the spectra while removing the grid structure and noise simultaneously. It has the potential to be extended to all spectroscopic measurements to eliminate other extrinsic signals and enhance the spectral quality based on the self-correlation of the spectra solely.
    
\end{abstract}

\maketitle
\section{Introduction}\label{section1}
		
	In the past decades, ARPES has driven the research of novel quantum materials with its incredible ability to directly probe the electronic structures  \cite{Damascelli2003,Hashimoto2014,Yang2018,Zhang2020,Lv2019,Lv2021,Sobota2021}. Owing to the rapid development of experimental techniques, highly rapid data acquisition modes ("fixed" or "dithered" modes) are frequently required in many application scenarios. For example, the fast scanning mode is often used in spatial-resolved ARPES experiments to map the energy band spectra of a wide area of the sample \cite{Iwasawa2017,Cattelan2018}. It is also preferred when the spot size of the laser beam is tiny, in which case the measurement should be limited to a very short time to avoid sample damage \cite{Grass2010}. In addition, the fast scanning mode can save a lot of acquisition time for measuring higher-dimensional data, such as band structures in two-dimensional momentum space or dynamical electronic structures in time-resolved ARPES \cite{Orenstein2012,Aoki2014,Smallwood2016,Bao2021,Torre2021}.
    
\begin{figure*}[t]
	\centering
	\includegraphics[width=0.99\textwidth]{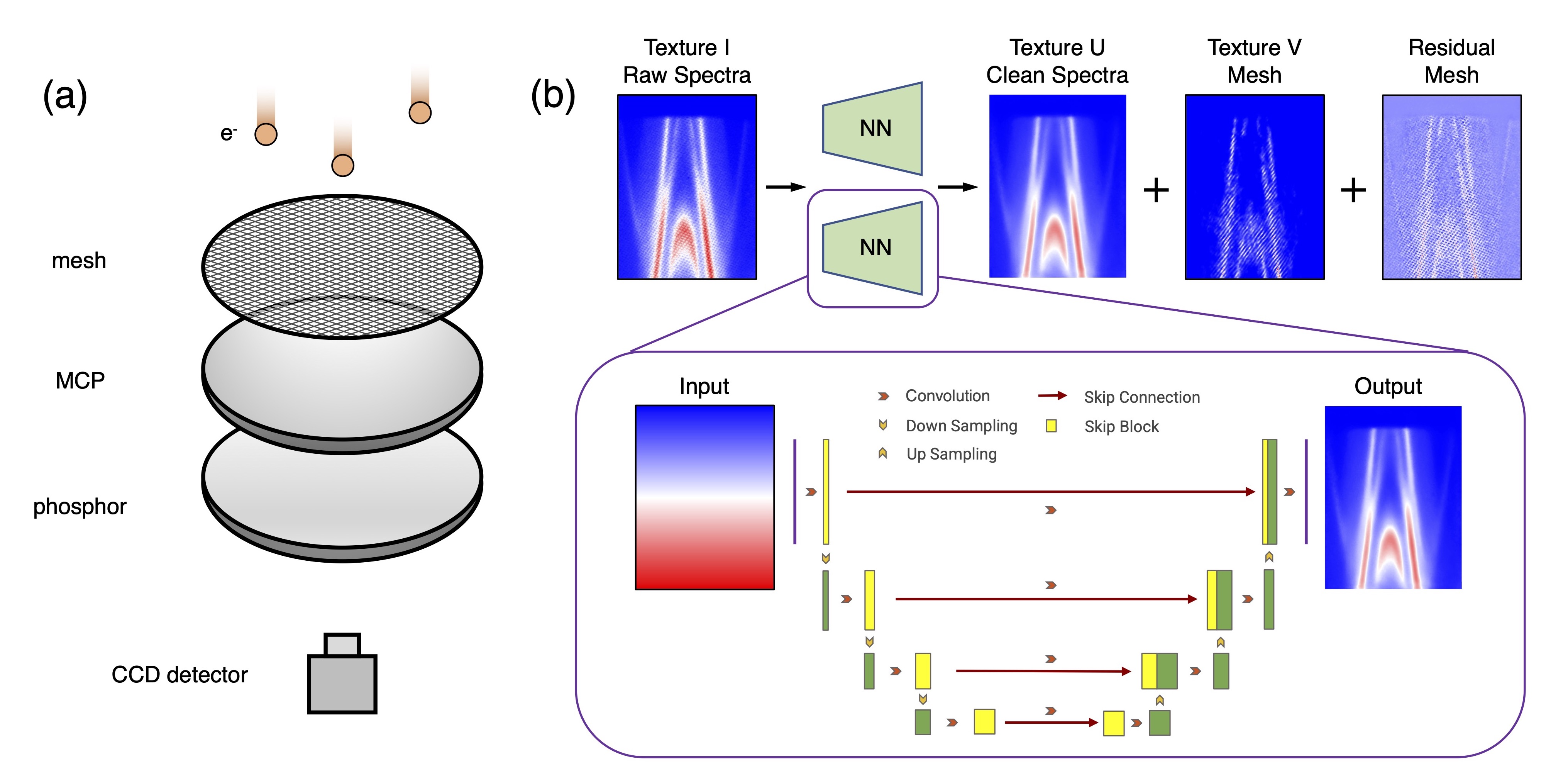}
	\caption{(a) An illustration of the ARPES analyzer. (b) The raw spectra image is parameterized as a mixture of two textures represented by two individual neural networks of the same structure. The first texture aims to recover the spectra and the other finds the grid structure somewhat attached to the energy bands. The residual is the general grid structure since the neural network does not learn all the features in the input image.}
	\label{Fig. method}
    \end{figure*}

    However, because of the metal mesh in front of the analyzer CCD, the spectra obtained using the fast scanning mode typically have a grid-like structure as shown in Fig. \ref{Fig. method}(a) \cite{Sobota2021,Liu2021}, which hinders the direct observation of energy band features. Although the grid structure may be averaged out using the swept mode, measurements in this mode often cover a large portion of unwanted energy ranges, causing a significant time waste in opposition to the original purpose. Post-spectral processing techniques and methods are hence needed to remove this grid structure. 
    
    A traditional method is to use Fourier filtering by converting the real space spectra to the Fourier domain \cite{Liu2021}. But simply erasing the peaks in Fourier space may lose intrinsic information because the grid structure is not strictly periodic and linearly superimposed. For instance, the peaks and energy bands in Fourier space may merge together when their widths are comparable, which makes it difficult to remove the peaks without losing information about the energy bands. It may become even more challenging when the data quality is noisy or not good enough so that the peaks in Fourier space cannot be well identified. Therefore, the traditional Fourier filtering method requires the peaks in Fourier space to be sharp and distinguishable from the intrinsic bands, which essentially limits the range of its application scenarios.
    
    Fortunately, machine learning methods have shown strong capabilities in spectra processing \cite{Zhang2011,He2017,Peng2020,Kim2021}. Deep learning-based methods can achieve better performance than traditional mathematical Gaussian smoothing methods in removing noise from spectra \cite{Kim2021}. More surprisingly, we have shown that a noisy spectra image can be decomposed into an image of clean spectra and an image corresponding to noise, provided that the noise is sparse and not so coherent with the intrinsic energy bands. This inspired us to view the grid structure as an extrinsic signal, so that the spectra may be decomposed into a clean part and a grid part.

    Following the above line of thought, we propose a deep learning-based method to remove the grid structure and identify the intrinsic signal in the measured ARPES spectra in this work. Our method utilizes the self-correlation information of the spectra themselves and parameterizes the observed ARPES data by two convolutional neural networks (CNNs). One network is designed to extract the clean spectra, and the other aims to extract the grid texture. We show that this enables us to preserve the signal of the energy bands in the spectra even if the grid width and energy bandwidth are comparable or the spectral quality is not so good. As a result, our method can nicely remove the grid structure and extend the application of the fast scanning mode, which is essential for ARPES experiments. Our idea can be applied to any extrinsic structures from other multidimensional data, thus improving the quality of the spectra obtained in more general scenarios.

\section{Methods}\label{sec:2}
    \begin{figure*}[t]
	\centering
	\includegraphics[width=0.8\textwidth]{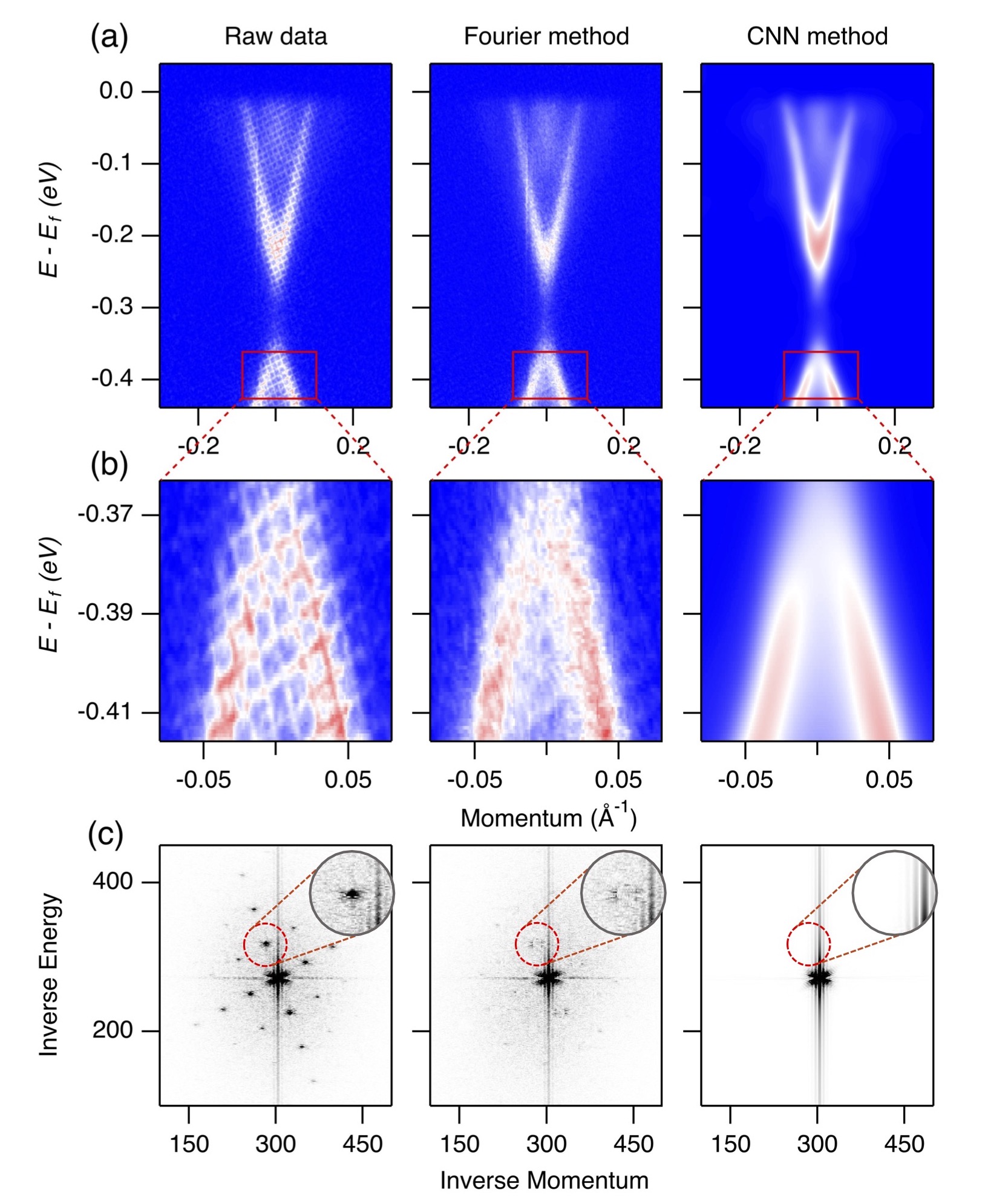}
	\caption{Comparison of the de-gridded results using different methods. (a) $\rm{MnBi_2Te_4}$ ARPES data along the cut through $\bar{\Gamma}$ and the de-gridded results based on the Fourier filtering method and our deep-learning method. (b) Close-up showing of the area in the red box region indicated in panel (a). (c) Corresponding Fourier transform results of panel (a). The inset shows the close-up of the first-order peak and energy band signals indicated with red circle.}
	\label{Fig. result}
    \end{figure*}
    
    We start by assuming that a complex image can be represented by a mixture of two or more simple components or textures, each of which can be viewed as an image. At first glance, this seems impossible and under-determined because the unknown parameters are more than the known ones. But this problem has been proved solvable in the high-dimensional analysis if one component is incoherent with another \cite{Candes2011, Wright2010}. In other words, pixels or patches are correlated within each texture but independent of those within another texture. This insight has led to successful algorithms showing good performance in computer vision \cite{Ulyanov2020,Gandelsman2019} and ARPES de-noising \cite{Huang2022} even without requiring a training set. 
    
    Motivated by these successes, we extend the above idea and develop an algorithm for de-gridding the ARPES data. As shown in Fig. \ref{Fig. method}(b), we attempt to decompose a raw spectra image $I$ into two textures $U$ and $V$ and expect that $U$ gives the clean spectra and $V$ approximates the grid structure. All the images $I,U,V$ may be viewed as matrices, and the entries in $U$ or $V$ should be correlated. We parameterize each texture via a U-shape neural network \cite{Ronneberger2015}, which has shown good performance in modeling correlated data and real-world tasks in computer vision \cite{Ronneberger2015}. $U$ and $V$ are then the outputs of two individual neural networks of the same structure elaborated in Appendix B.

    The summation of both textures $U,V$ should give the raw spectra $I$, which we take as the loss function. This yields
    \begin{equation}
        \min_{\theta,\phi}\|I-U_\theta - V_\phi \|_F^2,
        \label{Eq:loss function}
    \end{equation}
    where $\theta,\phi$ are the parameter collections for each neural network, and $\|. \|_F^2$ denotes the square of the Frobenius norm, which is defined as the summation of all entries of a matrix. The parameters are then optimized by stochastic gradient descent (SGD) with respect to the loss function given in eq. \eqref{Eq:loss function}. Our algorithm is implemented via Pytorch \cite{PyTorch} and the detailed training hyperparameters are listed in Appendix B.

    \textit{Noisy data.} In practice, noise is unavoidable in the raw data. To make the algorithm more practical, we combine de-noising techniques into this framework. The noise $s$ is assumed to be sparse because an image of measured spectra should not be corrupted everywhere. Following similar techniques \cite{You2020}, we decompose the raw spectra into three textures: the clean spectra $U$, the grid structure $V$, and the noise $s$. As discussed in \cite{Huang2022}, the noise can be further parameterized as $s= a\circ a - b\circ b$, where $a,b$ are two vectors with learnable entries to be optimized, and the symbol $\circ$ denotes their element-wise product. The loss function now becomes
    \begin{equation}
        \min_{\theta,\phi, a, b}\|I-U_\theta - V_\phi - a\circ a + b\circ b\|_2^2,
        \label{Eq:loss function-noisy}
    \end{equation}
    where every matrix is flattened as a vector and $\|\cdot\|^2_2$ denotes the square of $\ell^2$ norm. We find quite stable training with the form in Eq. \eqref{Eq:loss function-noisy}.

\section{Applications}\label{sec:3}
\begin{figure*}[t]
	\centering
	\includegraphics[width=0.99\textwidth]{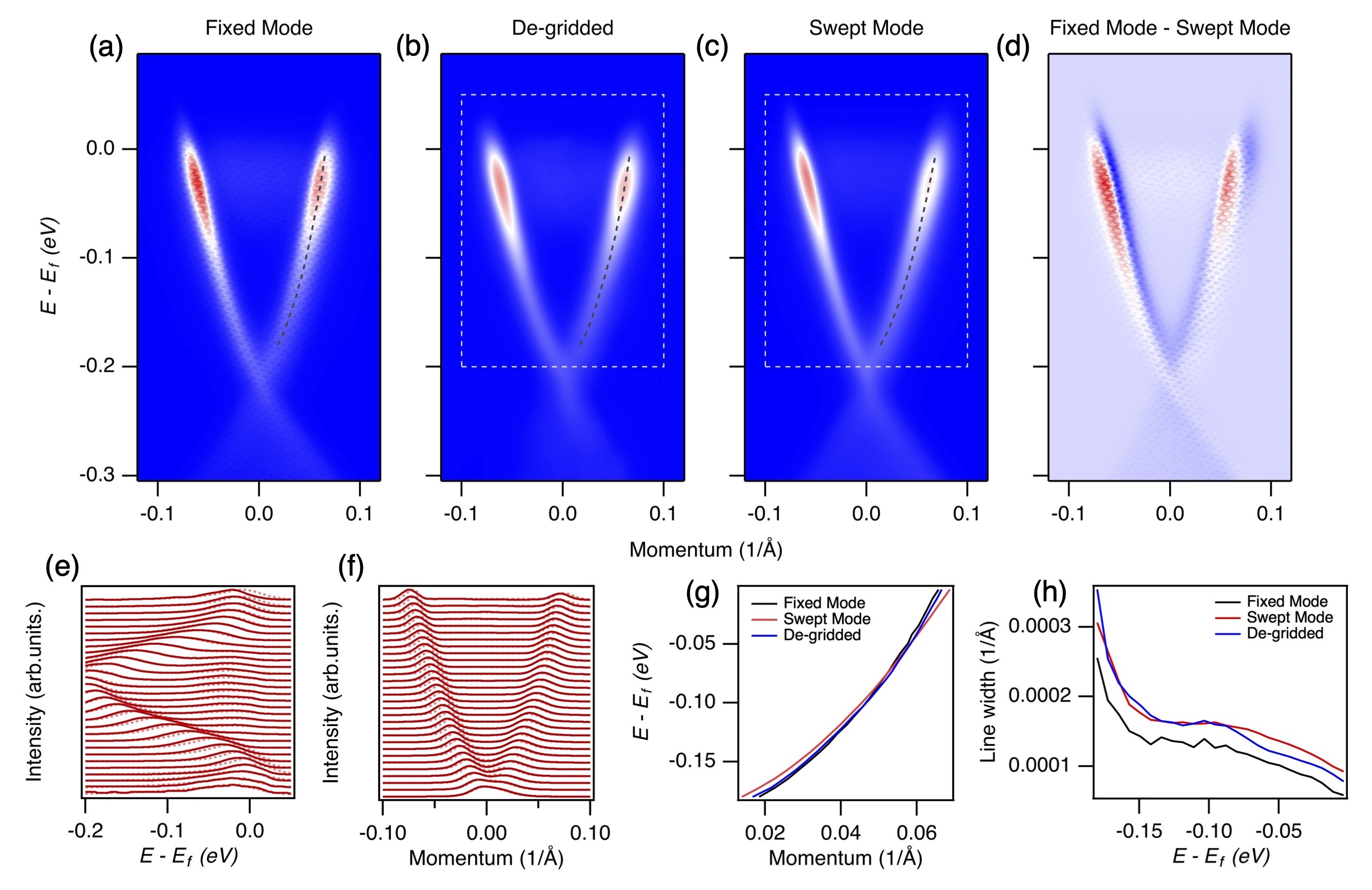}
	\caption{Quantitative analysis of the de-gridded results. (a) $\rm{Bi_2Se_3}$ ARPES data along the cut through ${\Gamma}$ measured with fixed mode. (b) The corresponding de-gridded results of panel (a) based on our deep-learning method. (c) Same as panel (a) but measured with swept mode. (d) Differential spectra of panel (a) and panel (b). (e) Energy distribution curves (EDCs) corresponding to the white box region indicated in panels (b) and (c). The dark red solid lines represent the de-gridded results and the light red dashed lines represent the swept mode results. (f) Corresponding momentum distribution curves (MDCs). (g) Peak position obtained by fitting the MDCs curve. The band dispersions of fixed mode spectra, de-gridded spectra, and swept mode spectra are indicated by grey dashed lines in panels (a), (b) and (c), respectively. (h) Peak width obtained by fitting the MDCs curves of fixed mode spectra, de-gridded spectra and swept mode spectra.}
	\label{Fig. result2}
    \end{figure*}	
    
    Fig. \ref{Fig. result}(a) shows the raw data and the de-gridded results using different methods for ARPES data of $\rm{MnBi_2Te_4}$ along the cut through $\bar{\Gamma}$. The raw data were measured in the fast scanning mode. The quasi-periodic grid structure is clearly visible in the raw data, as may be best viewed in the enlarged plot Fig. \ref{Fig. result}(b). The grid structures overlap with the energy band features, which may hinder the observation and analysis of some fine details. 
    
    Fig. \ref{Fig. result}(c) shows the spectra in the Fourier space after applying the Fourier transform to the data in Fig. \ref{Fig. result}(a). We see that the energy band signals are concentrated in the central region and along the lines due to its non-periodic nature, while the signals of the quasi-periodic grid structure yield the Fourier peaks distributed symmetrically near the diagonals of the image, whose locations represent the reciprocal vectors of the grid. The inset (red circle) shows the close-up of the first-order Fourier peak and energy band signals. They are quite close to each other, making it extremely difficult to erase the grid structure without losing intrinsic information.
    
    The de-gridded results by using the Fourier filter-based method and the deep learning-based method are compared in Fig. \ref{Fig. result}(b). While both methods can remove the grid structure from the original spectra, the quality of the resulting spectra processed by the deep learning-based method is obviously much better. We see that the deep learning-based spectra are very clean and smooth, but the Fourier filter-based ones are still rough and even contain some empty spaces. This is because the existence of the grid structure interferes with the acquisition of the intrinsic energy band signals in the measurement process. As a result, simply erasing the Fourier peaks fails to fill the missing intrinsic signal affected by the grid structure and thus loses the corresponding energy band features. Indeed, as shown in Fig. \ref{Fig. result}(c), it is difficult for the Fourier filter-based method to completely remove the Fourier peaks without affecting the intrinsic energy band signal. By contrast, deep learning-based methods can naturally and directly extract the complete intrinsic energy band signal through local correlation without losing critical information. It not only eliminates the Fourier peaks entirely but also completes the intrinsic energy band signal, making the spectra quite smooth and comprehensive. 

    To evaluate the fidelity and reliability of our deep-learning method, we conducted quantitative analysis on the ARPES data of the topological insulator $\rm{Bi_2Se_3}$ along the cut through ${\Gamma}$. As we can see, the data measured under fixed mode in Fig. \ref{Fig. result2}(a) show clear grid structures, while those under swept mode in Fig. \ref{Fig. result2}(c) only show clean energy band information. A straightforward comparison shows that our de-gridded results in Fig. \ref{Fig. result2}(b) have a good performance and show very comparable energy band structures, indicating that the convolutional neural network method removes the grid structure while retaining the intrinsic energy band information to a great extent. It should be noted that our method assumes a linear superposition of the energy band and grid structure in the spectra. From the differential spectra of fixed mode and swept mode shown in Fig. \ref{Fig. result2}(d), the extracted part contains both features of the grid structure and energy band, indicating that the spectra are not exactly their linear superposition. Nevertheless, the naive linear superposition assumption may not be a bottleneck in practice. Its good performance indicates that the correlation between grid structure and energy band is weak so the correlation information can be assembled into the extracted grid texture under the linear superposition assumption.

    To further show the good performance of our deep-learning method and the linear superposition assumption, we have also conducted quantitative line shape analysis. Figs. \ref{Fig. result2}(e) and \ref{Fig. result2}(f) compare the energy distribution curves (EDC) and momentum distribution curves (MDC) of the de-gridded data and the swept mode data in the white box region indicated in Figs. \ref{Fig. result2}(b) and \ref{Fig. result2}(c). It can be seen that the curves of the de-gridded data (dark red solid lines) and the swept mode data (light red dashed lines) are more closely matched, indicating that the deep-learning method can yield comparable results with the swept mode, which preserves clean energy band signal to a large extent without losing intrinsic information.
    
    In addition, since important physics often occurs in the energy bands near the Fermi level, we have made further analysis of band features near the Fermi level to show the performance of our algorithm. Specifically, we extracted the peak position and width of the energy bands near the Fermi level region by fitting the MDCs. As compared in Figs. \ref{Fig. result2}(g) and \ref{Fig. result2}(h), the fitting results from the fixed mode, de-gridded, and swept mode data are almost the same, demonstrating that our deep-learning method preserves the intrinsic information of the energy bands near the Fermi level region. Note that the peak widths of the fixed mode data are smaller, probably because the grid structure enhances the local intensity and reduces the peak widths. More examples are given in Appendix A to illustrate the universality and robustness of our de-gridded method.

    However, there are still some scenarios in which our method may not be as applicable. Based on our assumption of linear superposition, the energy band signal is required to occupy a significant proportion of the spectra so that its correlation information is sufficient for the algorithm to extract. For example, when the acquisition time is too short, the signal-to-noise ratio of the spectra may be too low for the algorithm to distinguish genuine signals from spurious signals such as grid and noise. Moreover, for some extremely fine structures, the presence of a grid may obscure a significant amount of intrinsic information such that the algorithm cannot accurately reconstruct the signals. Additionally, since our method relies on the assumption of sparse noise, our algorithm is not strictly applicable when the noise is particularly dense or when there are bad channels in the spectra. Therefore, while our approach has demonstrated good performance in many scenarios, it is important to be aware of these limitations and to evaluate its applicability carefully in each specific case.

\section{Discussion}\label{sec:4}
	\subsection{The training dynamics}
    
    As described above, the deep learning-based method can remove the weak-correlated extrinsic grid structure and extract the strong-correlated intrinsic energy band signals. The reason for its good performance may be ascribed to the correlation of neighboring pixel values in the spectra of ARPES experiments. Specifically, because of the finite  linewidth of the energy band, the regions of band features generally span a finite range and contain a number of MCP events. Since a single event on the MCP is a spot that occupies multiple pixels in the CCD detector, the energy band usually covers many CCD pixels so that the intensity of neighboring CCD pixels does not exhibit large fluctuations. By contrast, the size of the grid structure in most cases is much smaller than the linewidth of the energy band and contains fewer MCP events so that the changes of CCD pixel values brought by the grid structure are more localized and their global correlation is much weaker. Therefore, even if the grid has contaminated the information at several CCD pixels, the value of those CCD pixels can be retrieved by inferring the most statistically probable value from the CCD pixel values in a larger area nearby while retaining the intrinsic information from the energy band.

\begin{figure}[H]
    \centering
    \includegraphics[width=0.5\textwidth]{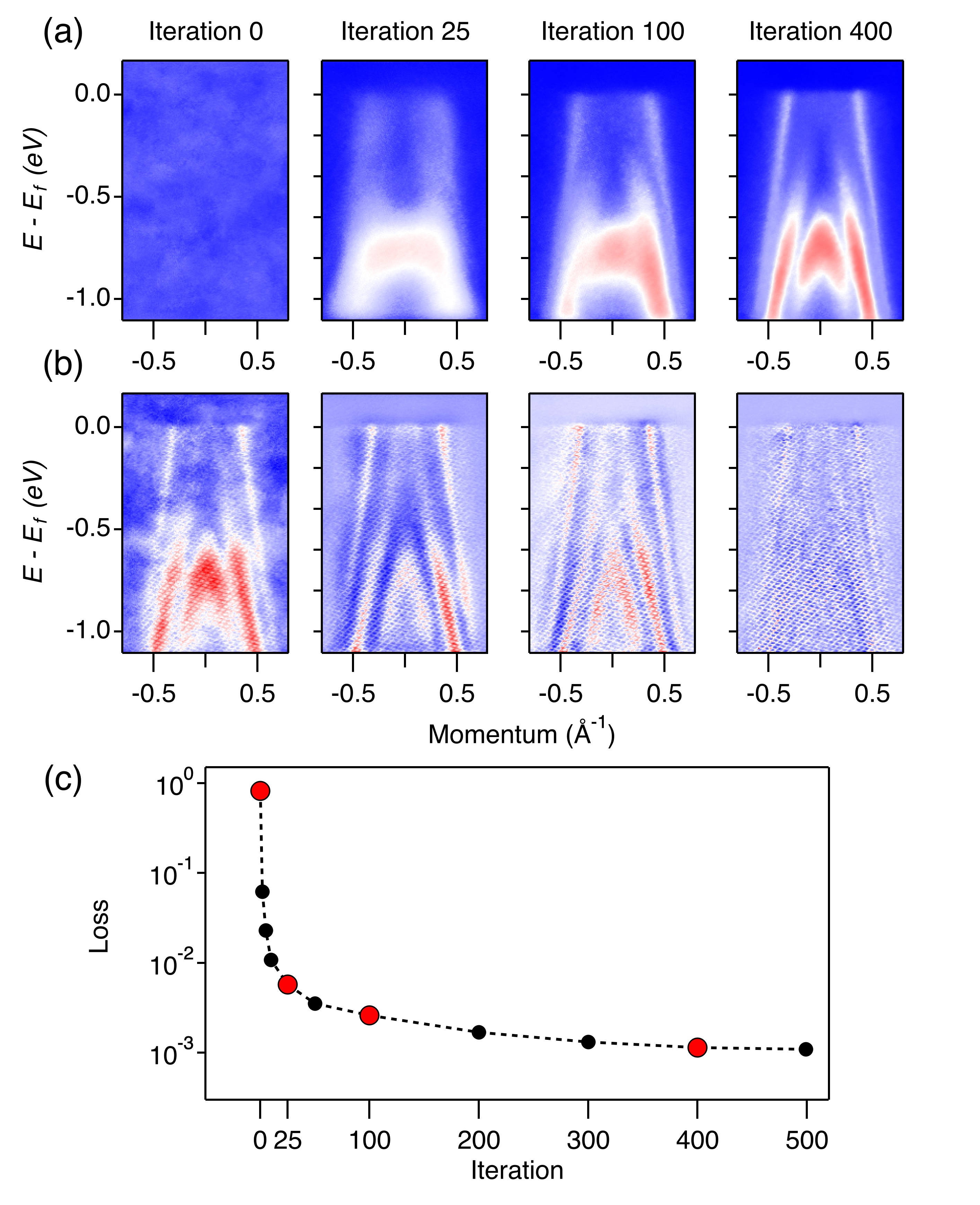}
    \caption{Intermediate results during the training process depending on iteration. (a) $\rm{Ir_{1-x}Pt_xTe_2}$ ARPES data and iteration-dependent de-gridded results for the iteration steps 0, 25, 100, and 400, which denotes the number of loops in the computing process. (b) Corresponding residual spectra of (a). (c) Loss $\rm{L(x,f(x'))}$ as a function of the iteration. The corresponding iteration numbers plotted in (a) and (b) are highlighted with enlarged red circles.}
    \label{Fig. discussion1}
    \end{figure}

    To demonstrate how the algorithm extracts the intrinsic energy band signal, we present the dynamical processing results of $\rm{Ir_{1-x}Pt_xTe_2}$ ARPES data in Fig. \ref{Fig. discussion1}. It can be seen that as the iteration of the computing process increases, the intrinsic energy band signals are gradually extracted [Fig. \ref{Fig. discussion1}(a)], leaving only the grid structure [Fig. \ref{Fig. discussion1}(b)]. Specifically, at zero iteration, the extracted signal is the random initial input. As the number of iterations increases to 25, the contours of the two energy bands start to emerge. Although the band structures are very crude at this stage, they have already covered all the areas where the real energy bands are located. As the iteration increases to 100 and 400, more detailed structures are revealed and we eventually obtain the clear spectra. The computing process tends to extract the signals [Fig. \ref{Fig. discussion1}(a)] with more local and stronger correlations from the remaining spectra [Fig. \ref{Fig. discussion1}(b)] to produce the intrinsic band structures. 

    Fig. \ref{Fig. discussion1}(c) plots the evolution of the loss function $L(x,f(x'))$ with the training iteration. We see that it decreases rapidly from the initial point to iteration 25. As shown in Fig. \ref{Fig. discussion1}(a), the algorithm first captures the general shape of the energy bands, indicating that the weak global correlations can be extracted quite easily. As the correlation information in the energy bands to be extracted becomes stronger and more local, the loss decreases more slowly. After approximately 300 iterations, the loss converges to a small value. The  convergence shows that the algorithm can no longer extract more local correlation information and the model is not overfitted to the grid structure, thus ensuring that the energy band information of the intrinsic signals may be largely preserved. From this point of view, we may conclude that the neural network first learns low-frequency long-range features and then adds high-frequency local details during the training.

\subsection{Effect of de-noising}

\begin{figure}[H]
        \centering
        \includegraphics[width=0.5\textwidth]{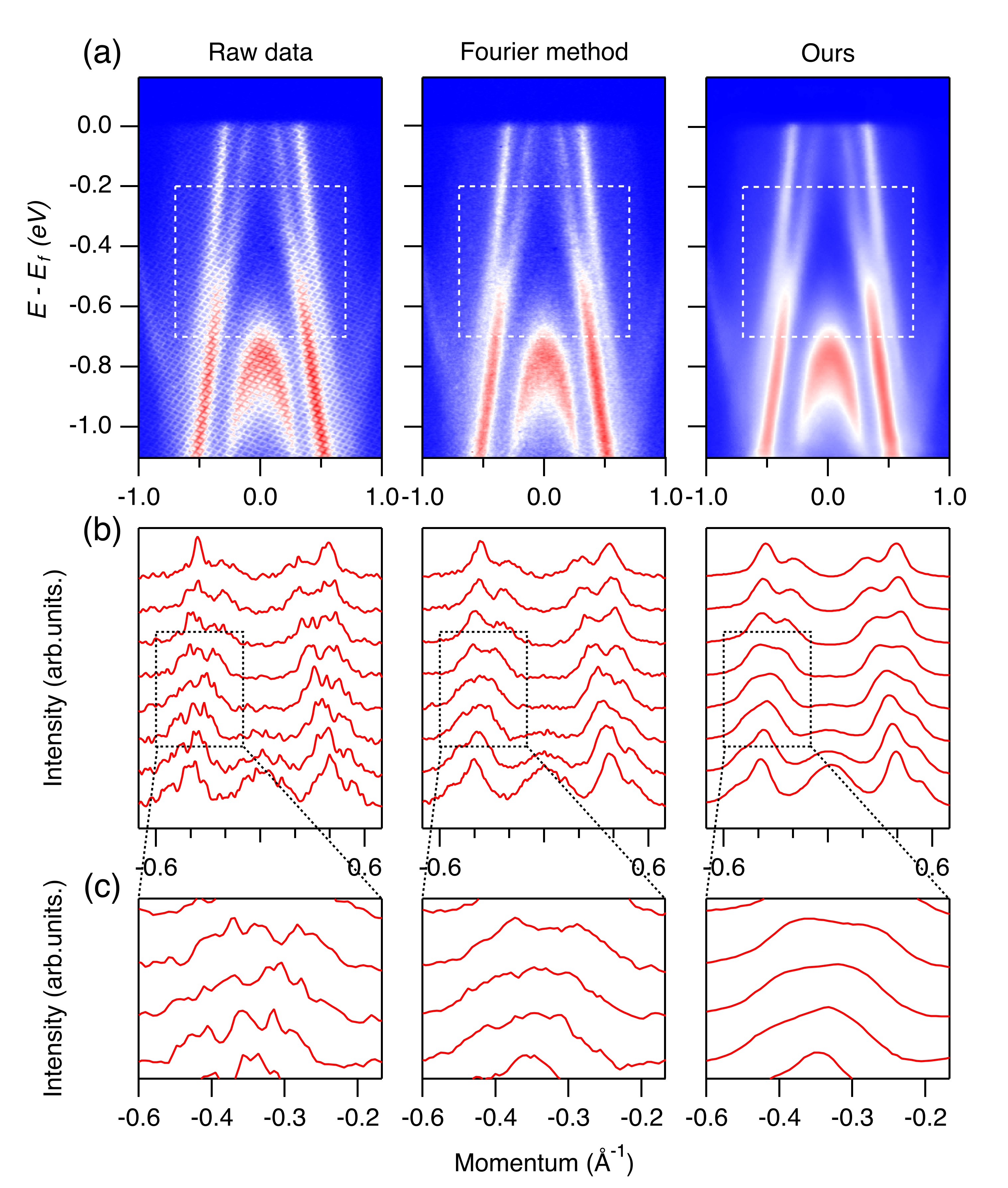}
        \caption{Comparison analysis of different methods. (a) The raw data of ARPES spectra for $\rm{Ir_{1-x}Pt_xTe_2}$ and the de-gridded results based on the Fourier filtering method and the deep-learning method. (b) Corresponding momentum distribution curves (MDCs) of the area in the white box region indicated in panel (a). (c) Close-up showing of the area in the black box region indicated in panel (b).}
        \label{Fig. discussion2}
    \end{figure}

    As the grid structure is removed, the spurious noise is also seen to be removed from the spectra and the quality of the resulting spectra is greatly improved. To quantitatively evaluate the performance of de-noising, we use line shape analysis on $\rm{Ir_{1-x}Pt_xTe_2}$ ARPES data. As shown in Fig. \ref{Fig. discussion2}(a), our method can preserve the intrinsic band structures while removing the noise compared to the original data and the de-gridded results using the Fourier filtering method. This is more clearly seen in the smoother momentum distribution curves (MDCs) shown in Figs. \ref{Fig. discussion2}(b) and \ref{Fig. discussion2}(c) for the area indicated by the white box region in Fig. \ref{Fig. discussion2}(a).   

    The reason why the noise can be simultaneously removed should be attributed to the foundation of our method, which respects the locally correlated intrinsic energy bands in noisy and grid-like spectra.  It has been proved \cite{Candes2011} that if a signal is a superposition of some incoherent components, each component can be identified by proper parameterization and optimization. In the analysis of spectroscopic measurements, both the spurious noise signals and grid structures are weakly correlated with the intrinsic energy bands. The weak correlation of the noise may be understood since human eyes are able to identify the intrinsic energy band from a noisy measurement, thus indicating that the noise does not overlap the full information of the intrinsic energy band. The superposition of the grid and noise, combined with the intrinsic energy bands, gives the raw spectra. We may hence view them as three nearly independent parts. Thus, by parameterizing them  via three corresponding neural networks in eq. \eqref{Eq:loss function-noisy}, we can separate them from the total spectra and identify the intrinsic energy bands as shown in Fig. \ref{Fig. discussion2}(a). This setting is easily generalizable, so our method can be extended to many application scenarios to remove different extrinsic signals simultaneously and achieve clear and desired spectra. Although our algorithm does not assume Poisson statistics about the noise. We have tested the robustness of our algorithm in removing Poisson distribution noise and found that its variance is consistently minimal across different spectra. This indicates that our algorithm is highly reliable and does not introduce additional errors in the quantitative analysis, thus providing a robust method for improving the quality of the spectra. In addition, We mention that it might be difficult to extract error bars when analyzing the de-noised data because we do not explicitly deal with the Poission statistics.
    
\section{Conclusion}\label{sec:5}
	In this work, we propose a deep-learning-based method to eliminate the grid-like structure in fast scanning mode for ARPES spectroscopy. This is quite helpful for saving measurement time and enhancing spectral quality. Compared to the traditional Fourier filtering methods, our deep learning method can extract intrinsic information directly by making use of the local correlation nature of the energy band signals. It can remove extrinsic features including the grid structure and noise simultaneously, and thus greatly improve the spectra. Our algorithm may also be extended to other spectroscopic measurements for removing unwanted signals, including quasi-periodic structures, spurious signals, noise, etc.
 
	\begin{figure*}[t]
        \centering
        \includegraphics[width=0.90\textwidth]{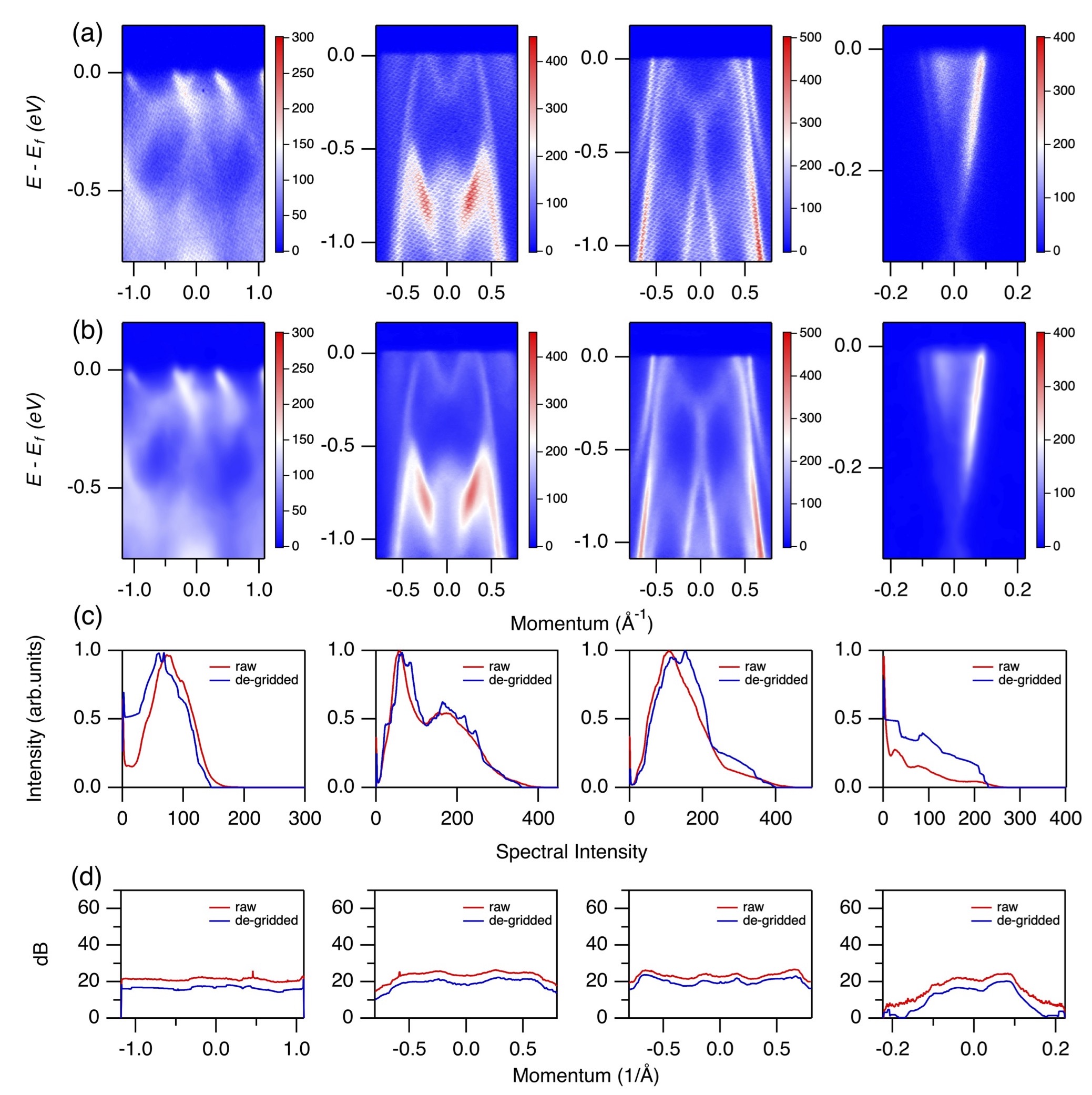}
        \caption{The de-gridded results for more ARPES spectra. (a) The raw data with grid-like structure. (b) The corresponding de-gridded results based on our deep-learning method. (c) The distribution on the spectral intensity of the raw data and de-gridded data. (d) The dynamic range of EDCs for different momentum positions based on the raw data and de-gridded data.}
        \label{Fig. Appendix1}
    \end{figure*}
    
\section*{ACKNOWLEDGEMENTS}
We thank M.-J Pan, F.-M. Chen, J.-R. Huang, and B. Jiang for useful discussions, and Z.-C. Rao, S.-Y. Gao, and W.-H. Fan for data support. This work was supported by the National Natural Science Foundation of China (NSFC Grants Nos. U1832202, 11888101, 11974397, 12174429), the Chinese Academy of Sciences (Grant Nos. QYZDB-SSW-SLH043, XDB33000000, and XDB28000000), the Informatization Plan of Chinese Academy of Sciences (Grant No. CAS-WX2021SF-0102), and the Synergetic Extreme Condition User Facility (SECUF).

J.L. and D.H. contributed equally to this work.

\section*{APPENDIX A: MORE EXAMPLES}
    Figs. \ref{Fig. Appendix1}(a) and \ref{Fig. Appendix1}(b) show more de-gridded examples based on the deep-learning method. Each de-gridded spectrum is normalized according to the total intensity of its raw spectrum, while the range of colorbar is set to the same.  Regardless of sharp or diffuse energy bands in the spectra, our method can always remove the grid structure and extract well the intrinsic energy band information. This indicates that our method based on CNNs is quite stable and has a strong potential for  generalization.

    Fig. \ref{Fig. Appendix1}(c) shows the distribution of the spectral intensity obtained from Figs. \ref{Fig. Appendix1}(a) and \ref{Fig. Appendix1}(b), which can represent the dynamic range of the spectra. The distribution based on the de-gridded spectra covers most of that of the raw spectra, indicating that our de-gridding method sets no artificial limit on the extracted spectra. However, it is worth noting that the maximum spectral intensity in each distribution of the de-gridded data is smaller than that in the distribution of the raw data. This is because the intensity is enhanced by the grid structure in the raw data, so the maximum spectral intensity becomes lower when the grid is removed.

    To verify the effect of the de-gridded method more clearly, we calculate the dynamic range of each EDC curve at different momentum positions of the raw data and the de-gridded data according to the formula in dB:
    \begin{equation}
        dB=10\times\log_{10}\left(\frac{I_{\max}}{1+I_{\min}}\right)
        \label{Eq:dynamic range}
    \end{equation}
    where $I_{\max}$ and $I_{\min}$ denote the maximum and minimum intensity of each EDC curve, respectively. The results are compared in Fig. \ref{Fig. Appendix1}(d). The dynamic range of the de-gridded data is not much different from that of the raw data, indicating that our de-gridding method does not have a significant effect on the resolution of the spectra and the de-gridded results are acceptable. The dynamic range after de-gridding is smaller at different momentum positions, probably because of the removal of the grid structure signal, which reduces the maximum intensity and hence the dynamic range.

\section*{APPENDIX B: IMPLEMENTATION DETAILS}
    
    The neural network parameters are listed in Table \ref{Tab:arch}. We use batch normalization (BN) \cite{Ioffe2015} every layer, and set the parameter of leaky-ReLU to $0.2$. This algorithm is not sensitive to the input, so we use similar inputs as reported in previous work \cite{Ulyanov2020,You2020}, plus a small fraction ($\sim 0.01$) of the raw data in this work. To achieve better results, discrepant learning rates should be used for the three neural networks. We find $(0.2, 2, 600)$ is an appropriate parameter choice for many ARPES images.

\begin{table}[ht]
\caption{Neural network architecture}
\label{Tab:arch}
\setlength\tabcolsep{7pt}
\begin{tabular}{ccc}
  \hline\noalign{\smallskip}
  Encoder network\\
  \noalign{\smallskip}\hline\noalign{\smallskip}
  Input spectra $x$ \\
  Conv2d, BN, $5 \times 5 \times 16$, stride=2, lReLU\\
  Conv2d, BN, $5 \times 5 \times 32$, stride=2, lReLU\\
  Conv2d, BN, $5 \times 5 \times 64$, stride=2, lReLU \\
  Conv2d, BN, $5 \times 5 \times 128$, stride=2, lReLU \\
  Conv2d, BN, $5 \times 5 \times 128$, stride=2, lReLU \\
  \noalign{\smallskip}\hline
  Decoder network\\
  \noalign{\smallskip}\hline
  Conv2d, BN, $5 \times 5 \times 128$, stride=1, lReLU \\
  Upsampling, Conv2d, BN, $5 \times 5 \times 128$, stride=1, lReLU \\
  Upsampling, Conv2d, BN, $5 \times 5 \times 64$, stride=1, lReLU \\   
  Upsampling, Conv2d, BN, $5 \times 5 \times 32$, stride=1, lReLU \\
  Upsampling, Conv2d, BN, $5 \times 5 \times 16$, stride=1, lReLU \\
  Conv2d, $1 \times 1 \times 1$, stride=1, sigmoid\\
  \noalign{\smallskip}\hline
\end{tabular}
\end{table}

\begin{figure*}[t]
        \centering
        \includegraphics[width=0.8\textwidth]{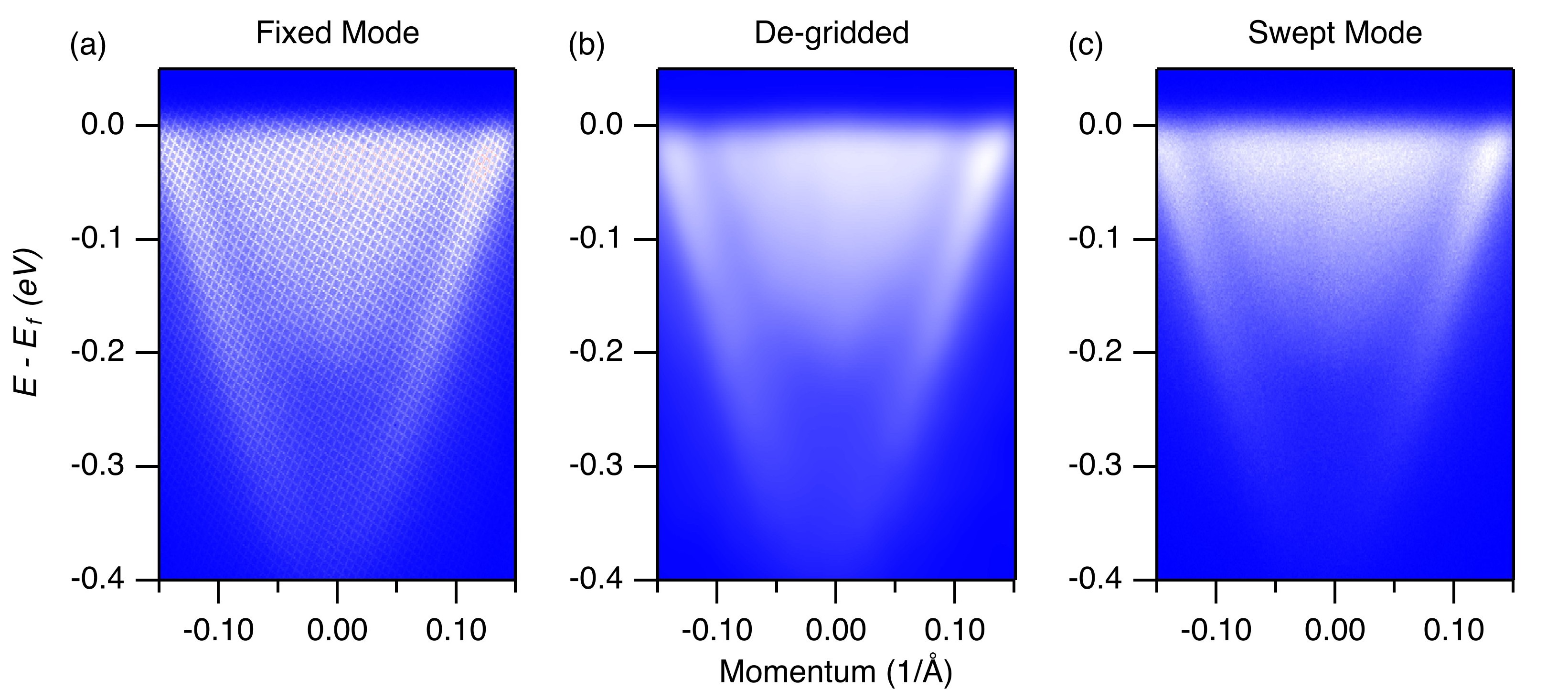}
        \caption{De-gridded results of pulse-counting mode spectra. (a) The pulse-counting mode ARPES spectra of $\rm{Bi_2Se_3}$ under fixed mode. (b) Corresponding de-gridded results of (a) based on our deep-learning method. (c) The pulse-counting mode ARPES spectra of $\rm{Bi_2Se_3}$ under swept mode.}
        \label{Fig. Appendix2}
    \end{figure*}
    
\section*{APPENDIX C: Pulse-counting mode spectra}

    To verify the generalizability of our algorithm, we applied the de-gridding method to the ARPES spectra of $\rm{Bi_2Se_3}$ under pulse-counting mode. In this mode, an MCP event will not occupy several CCD pixels, but only a single CCD pixel, i.e. the CCD pixels are independent of each other. Fig. \ref{Fig. Appendix2} shows the de-gridded results of pulse-counting mode spectra, where each spectrum is normalized according to the total intensity.
    
    Even though the signal intensities are independent between CCD pixels, the de-gridded results in Fig. \ref{Fig. Appendix2} show that our algorithm can still remove the grid signal nicely and yield comparable results as the swept mode. This indicates that although the CCD pixels are acquired independently, the energy band has a certain linewidth that can occupy a finite range of CCD pixels, so the algorithm can use this correlation to extract clean energy band information. This suggests that our algorithm only depends on the correlation between energy band signals and hence may be universally applied in different modes of ARPES applications.


\end{document}